\documentclass[%
 reprint,
 amsmath,amssymb,
 aps,
]{revtex4-2}
\usepackage{graphicx}
\usepackage{bm}
\usepackage[dvipsnames]{xcolor}
\usepackage{amsmath}
\usepackage{float}
\usepackage{caption} 
\usepackage{xcolor}
\usepackage{changes}

\begin{document}

\title{Efficient Gaussian Process Regression for prediction of molecular crystals harmonic free energies}

\author{Marcin Krynski}
\email{marcin.krynski@pw.edu.pl}
\affiliation{Fritz Haber Institute of the Max Planck Society, Faradayweg 4-6, 14195 Berlin, Germany}
\altaffiliation{Present address: Warsaw University of Technology, Faculty of Physics, Koszykowa 75, 00-662 Warsaw, Poland
}

\author{Mariana Rossi}
\affiliation{Fritz Haber Institute of the Max Planck Society, Faradayweg 4-6, 14195 Berlin, Germany}
\affiliation{MPI for the Structure and Dynamics of Matter, Luruper Chaussee 149, 22765 Hamburg, Germany}

\begin{abstract}
We present a method to accurately predict the Helmholtz harmonic free energies of molecular crystals in high-throughput settings.
This is achieved by devising a computationally efficient framework that employs a Gaussian Process Regression model based on local atomic environments.
The cost to train the model with \textit{ab initio} potentials is reduced by starting the optimisation of the framework parameters, as well as the training and validation sets, with an empirical potential.
This is then transferred to train the model based on density-functional theory potentials, including dispersion-corrections.
We benchmarked our framework on a set of 444 hydrocarbon crystal structures, comprising 38 polymorphs, and 406 crystal structures either measured in different conditions or derived from them.
Superior performance and high prediction accuracy, with mean absolute deviation below 0.04 kJ/mol/atom at 300 K is achieved by training on as little as 60 crystal structures.
Furthermore, we demonstrate the predictive efficiency and accuracy of the developed framework by successfully calculating the thermal lattice expansion of aromatic hydrocarbon crystals within the quasi-harmonic approximation, and predict how lattice expansion affects the polymorph stability ranking.
\end{abstract}

\maketitle

\section{Introduction}

Polymorphism and the prediction of the energetic stability of a crystal polymorph are a fundamental problem of condensed matter physics, especially for the research and applications of molecular crystals.
Polymorphism is the capability of solid materials to form more than one distinct crystal structure \cite{polimorph,C5CS00227C}.
It is particularly pronounced when multiple atomic or molecular packing arrangements are characterised by a similar free energy.
The physicochemical properties of these systems, such as mechanical and optical characteristics, melting point, chemical reactivity, solubility or stability are tied strongly to the crystal morphology, therefore increasing the relevance of a comprehensive structure screening and the prediction of the relative stability of polymorphs for a broad range of industries \cite{Davey2002}.

High-throughput computational screening of crystal structures based on free energies is rarely performed due to its high complexity as well as large computational effort, in particular if a first-principles potential energy surface is required \cite{Hojaeaau3338}.
It is more common to evaluate the relative stability of crystal polymorphs by calculating the lattice energy taking into account only potential energy contributions \cite{C5TC04172D, CURTAROLO2005155, PhysRevX.3.041035, C3CS60279F, C7SC04665K}, effectively disregarding enthalpic and entropic contributions at finite temperature \cite{Curtarolo2013, C5CS00227C}.
Finite pressure contributions when comparing different phases at different pressures is typically of a lower magnitude, reaching only about $1$ kJ/mol/molecule for pressure difference of several gigapascals.
It was shown \cite{C5CE00045A} that even if the vibrational free energy difference between two given polymporphs lies typically around $2$ kJ/mol/molecule, it is sufficient to cause a rearrangement of the polymporph relative stability ranking.
Furthermore, even when the vibrational contribution to the relative stability is taken into account in a number of cases, the effect of the thermal expansion of the crystal unit-cell on the free energy is most frequently omitted.
This is due to the, typically low impact of the thermal expansion on the free energy (around $1-2$ kJ/mol/molecule \cite{C6CP05447A}), which is, nevertheless, also sufficient to affect the polymporph stability ranking.

The vibrational part of the free energy can be accessed by, among others, two straightforward types of calculation: within the harmonic approximation given by lattice dynamics calculations \cite{latdyn, phd} and with statistical sampling methods that accounts for all anharmonic contributions, for example via thermodynamic integration (TI)  \cite{Vega_2008, PhysRevLett.94.145701, doi:10.1063/1.476566, PhysRevLett.117.115702, PhysRevB.97.054102}.
Even though methods like TI are more accurate, they are also extremely computationally demanding, requiring a large amount of statistical sampling in order to achieve the necessary accuracy.
This renders this technique often impossible to carry out within a high-throughput setting.
Approximations to the contribution of anharmonic terms to the free energy can be accessed by a number of other methods that are less computationally demanding.
However, such approximations have been shown not to present a significant improvement over the much less computationally demanding harmonic approximation for the investigation of polymporph relative stability \cite{Kapil2019}.
Still, harmonic lattice dynamics are not a viable solution for high-throughput screening if force evaluations are a bottleneck, since the calculation typically involves hundreds of force evaluations for a single structure (or costly perturbation theory techniques), considering the full unit cell. 

Within the last decades, the rapid increase of computer power, allied to the rise of machine learning (ML) and big-data algorithms in the realm of material science, allowed for large-scale screening of materials properties, including those related to polymorphism \cite{doi:10.1063/1.1458547, doi:10.1063/1.2210932, B719351C, Pickard_2011, doi:10.1080/0889311X.2010.517526, Curtarolo2013, PhysRevLett.107.015701, Oganov2019, PhysRevLett.120.145301}.
There are only a handful of examples where vibrational free energies \cite{Legrain2017}, or other
quantities related to the vibrational density of states \cite{Legrain2018VibrationalPO, PhysRevX.4.011019, PhysRevX.6.041061, Raimbault_2019}, were successfully predicted with the assistance of machine learning (ML) methods.
Those methods, however, do not focus on high transferability, or, if they do, rarely achieve the necessary accuracy to differentiate between polymorphs.
Clearly, if one could train a very accurate ML interatomic potential for a large class of systems, it would represent the best solution for the evaluation of lattice energies and free energies at the same time. 
However, despite the exceptional performance of many such potentials, typical root-mean-square errors on the forces  lie around 20 meV/\text{\AA}/atom \cite{george2020combining, PhysRevB.97.054303, PhysRevX.8.041048, C8CP05771K, Behler_2014, doi:10.1063/1.3095491}.
With such errors, the expected prediction accuracy of phonon modes is $\pm 0.15$ THz for the best performing potentials \cite{george2020combining}.
If the resulting phonon accuracy, as in \cite{george2020combining}, is assumed to be constant along the entire frequency range, the harmonic free energy calculation error amounts to $0.38$ kJ/mol/atom.

In this study, we target high accuracy and low computational cost for harmonic free energy predictions. We build a model for the prediction of Helmholtz harmonic free energies of molecular crystals based on Gaussian Process Regression (GPR) and Smooth Overlap of Atomic Positions (SOAP) \cite{C6CP00415F} descriptors for representing the local atomic environments.
We optimize the training and validation set selection with a computationally cheap empirical potential, confirm its transferability to a first-principles potential, and proceed to achieve a model with first-principles accuracy with a very low cost of training. 
For a set of hydrocarbon crystals, we are able to achieve a mean absolute error on the free energies of $0.04$ kJ/mol/atom.
We analyzed the stability ranking for a few families of hydrocarbon crystal polymorphs up to 300 K, highlighting the power and accuracy of the model.
Furthermore, this method can predict the anisotropic lattice expansion of these crystals, allowing a cheap evaluation of volume expansion and free energies in the quasi-harmonic approximation.

\section{Results and discussion}

Because it was shown \cite{Kapil2019} that the harmonic approximation to the free energy can be a suitable estimate for the computation of the relative stability between different structures of molecular crystals, this project focuses on predicting the harmonic Helmholtz free energies $F$.
Contributions from pressure that would be described instead by the Gibbs free energy are not considered because the structures regarded in this study are typically observed much below 1GPa of pressure, making this contribution to the free energy negligible.
Throughout this paper, for the sake of simplicity, $F$ is evaluated at the $\Gamma$ point of the Brillouin zone of a given unit cell. We consider unit cells larger than the primitive cell where needed (see Methods). The harmonic free energies are thus calculated as
\begin{equation}
    F(V, T) = \sum_{i=1}^{3N-3}
        \Big(
            \frac{\hbar\omega_i}{2} + k_BT\ln(1-e^{-\frac{\hbar\omega_i}{k_BT}})
        \Big),
\end{equation}
where $\omega_i$ is the frequency of a given phonon mode at the $\Gamma$ point.
When taking lattice expansion into account, the vibrational frequencies depend indirectly on the temperature such that $\omega_i = \omega_i(V(T))$.

\subsection{\label{sec:gpr}Definition of the GPR model}

The key assumption of the free energy prediction approach explored in this project is that even if free energies are defined only for the entire collection of atoms of the crystal structure, they can be decomposed into local contributions of atomic environments.
The approach of casting a global property on local environments was explored previously \cite{PhysRevLett.98.146401, PhysRevLett.104.136403} for the generation of an interatomic potential from quantum mechanical data.
The problem of the harmonic Helmholtz free energy prediction is approached by connecting the atomic-wise free energy to the full free energy by
\begin{equation}
    \bm{F}=\bm{M}^T\bm{f},
\end{equation}
where $\bm{F}$ is the vector with all measured free energies for a given crystal set of dimension $N_s$ (number of crystal structures in the training set), $\bm{M}$ is an incidence matrix of dimension $N_s \times N_{ae}$ (number of atom environments in the given set) and $\bm{f}$ is the vector of all, unobserved, atom-wise free energies in the chosen ensemble.
Then, the prediction of $\bm{f}$ in the training set is modeled as
\begin{equation}
    \bm{f}' = \bm{C} \bm{\alpha},\label{eq:atomwise}
\end{equation}
where $\bm{C}$ is the matrix containing the similarities between pairs of atomic environments (dimension $N_{ae} \times N_{ae}$), defined as
\begin{equation}
    C_{ij} = \sigma \text{e}^{-\displaystyle{\frac{\sum_{d=1}^{D}(q_{d,i} - q_{d,j})^2}{2 l^2}}}, \label{eq:kernel}
\end{equation}
where $\sigma$ is a scaling prefactor, and $\bm{q}_i$ is a vector of length $D$ describing local atomic environments.
$C_{ij}$ corresponds to the Gaussian kernel.
In Eq.~\ref{eq:atomwise},  $\bm{\alpha}$ is a vector of $N_{ae}$ weights for each atomic environment, such that
\begin{equation}
    \bm{F}' = \bm{M}^T \bm{C} \bm{\alpha}.
\end{equation}

Opening up this equation element-wise, the full free energy of one sample $i$ in the training set is given by
\begin{equation}
    F'_i = \sum_{j=1}^{N_{ae}} \sum_{k=1}^{N_{ae}} (\bm{M}^T)_{ij} C_{jk} \alpha_{k}.
    \label{eq:model}
\end{equation}

Optimizing the weights $\alpha_{k}$ is equivalent to minimizing the loss function
\begin{equation}
    L = \sum_i^{N_s} [F'_i - F_i]^2 + \sigma_{\epsilon}^2 \bm{\alpha}^T \bm{C} \bm{\alpha}
    \label{eq:loss}
\end{equation}
where $\sigma_{\epsilon}^2$ is a regularization parameter related to the variance of the noise of the data.

Finally, substituting Eq. \ref{eq:model} into \ref{eq:loss}, the minimization is straightforward and leads to
\begin{equation}
    \bm{\alpha} = \bm{M}(\bm{M}^T\bm{C}\bm{M}+\sigma_{\epsilon}^2\bm{I})^{-1} \bm{F},
\end{equation}
where $\bm{I}$ is the identity matrix of dimensions $N_{s} \times N_{s}$.
In this way, one can obtain the optimized weights with no need to define or observe atom-wise free energies. 

Finally the prediction of the free energy of a new structure that is not contained in the training set is achieved by calculating
\begin{equation}
    F(\bm{q}^*) = \sum_{i=1}^{N} \sum_{j=1}^{N_{ae}} (\bm{C}^{*T})_{ij} \alpha_j
\end{equation}
where $\bm{C}^{*}$ is the similarity matrix between the atomic environments $\bm{q}^*$ of the new structure to the ones in the training set, with elements
\begin{equation}
    C^*_{ij} = \sigma \text{e}^{-\displaystyle{\frac{\sum_{d=1}^{D}(q^*_{d,i} - q_{d,j})^2}{2 l^2}}}.
\end{equation}

All hyper-parameters for the GPR model and the representations were selected by minimising, using the steepest descent method, the negative log marginal likelihood function \cite{mlbook}
\begin{eqnarray}
    -\ln P(\bm{F}|(l, \sigma_\epsilon, \bm{\theta})) &= 
        \frac{1}{2}\ln|\bm{M}^T\bm{C}(l, \bm{\theta})\bm{M}+\sigma_{\epsilon}^{2}\bm{I}| + \nonumber\\
        &\frac{1}{2}(\bm{M}^T\bm{C}(l, \bm{\theta})\bm{M}+\sigma_{\epsilon}^{2}\bm{I})^{-1} + \nonumber\\
        &\frac{1}{2}\ln 2\pi
    \label{eq:mlh}
\end{eqnarray}
where $\bm{\theta}$ is a vector containing the hyperparameters of the representations entering $\bm{q}$.
The application of the steepest descent method is only guaranteed to find a local minimum. A wide space of hyper-parameters was considered in order to increase the probability of finding a global minimum.

In all supervised machine learning based models, the quality of the model strongly depends on the quality of the training set.
Typically, selecting the training set can be done by either a random selection of samples, given that the considered ensemble is fairly homogeneous, or by implementing methods that aim at covering the sampled domain by maximising the resulting prediction accuracy, such as the ``correlation'' clustering method \cite{C5SC04786B}, genetic optimization \cite{Browning2017} or $k$-fold cross-validation \cite{Hansen2013}.
Unfortunately, most of the methods from the latter group require a large pool of data for which the target property, like free energy in this case, is available.
In this study, because one of the objectives is to minimize the computational cost of obtaining a good training set, the applied procedure focuses on selecting an optimal training subset based exclusively on the geometrical parameters of the crystal structure.

For this purpose, the farthest point sampling (FPS) \cite{fps} method is applied, that searches for a subset  of the entire investigated crystal structure ensemble  that covers evenly all structural motifs of the sampled domain with minimal information overlap.
First, a similarity measures between molecular crystal structure $R_{a\rightarrow b}$ is defined according to the \textit{best-match structural kernel} \cite{Rupp2007} method, as it is needed for the application of the FPS
\begin{eqnarray}
    R_{a\rightarrow b} = \frac{1}{n_b}\sum_{i=1}^{n_{b}}\max_{j\in (1,n_{a})}C(\bm{q}_{j}^{a},\bm{q}_{i}^{b})
    \label{eq:Rab}
\end{eqnarray}
where $C(\bm{q}_{j}^{a},\bm{q}_{i}^{b})$ is the kernel matrix element defined in equation \ref{eq:kernel}, $\bm{q}_{j}^{a}$ and $\bm{q}_{i}^{b}$ are $i$-th and $j$-th atomic environment representations of structure $a$ and $b$ respectively, and similarly $n_a$ and $n_b$ are the number of atoms in those structures.
$R_{a\rightarrow b}$ defines how well atoms of structure $a$ can represent geometrical motifs of structure $b$ and  $R_{a\rightarrow b} \neq R_{b\rightarrow a}$.
In other words, it is possible that atomic environments of structure $a$ represent well those of structure $b$, while structure $b$ contains geometric features not present in $a$.
This method of defining the relationship between crystal structures is very similar to others typically chosen for such tasks \cite{PhysRevLett.112.083401, PhysRevLett.106.225502, doi:10.1063/1.4828704, Rupp2007, C6CP00415F}, with the difference that $R_{a\rightarrow b}$ is not invariant with respect to the crystal structure index ($R_{a\rightarrow b} \neq R_{b\rightarrow a}$) so it is not a similarity metric in a strict sense.
Next, according to the FPS algorithm, the training set is created by iteratively picking structures that are least represented by those already present in the training set.
Since any crystal structure can be used as the starting point for the FPS algorithm, the applied method selects $N_{\Omega}$ potential training sets, where $N_{\Omega}$ is the number of crystal structures in the considered ensemble.
In order to choose out of $N_{\Omega}$ potential training sets, we have investigated the scaled cumulative sum $I(N_{m})$ of the $R_{a\rightarrow b}$,
\begin{eqnarray}
    I(N_{m})= 
        \frac{
            \sum_{a}^{N_m}\sum_b^{N_{\Omega}}{R_{a\rightarrow b}}
        }
        {
            \sum_{a}^{N_{\Omega}}\sum_b^{N_{\Omega}}{R_{a\rightarrow b}}
        }
    \label{eq:information}
\end{eqnarray}
where $N_m$ is the total number of the molecular crystal configurations in the training set.
This quantity reveals how fast a given training set candidate converges to unity, which we consider to represent a  full coverage of the sampled feature space.
In another sense, the $I(N_{m})$ quantity can be seen as the description of the information acquisition during consecutive steps of the FPS algorithm.
Finally, training set with the highest recorded value of $I(N_{m})$ after all $N_{m}=60$ steps of the FPS algorithm is chosen.
The training set size of $N_{m}=60$ was chosen because above this number, the improvement of the prediction accuracy was too small to justify a larger training set and the associated increase in computational effort.

\begin{figure}[ht!]
    \centering
    \includegraphics[width=0.45\textwidth]{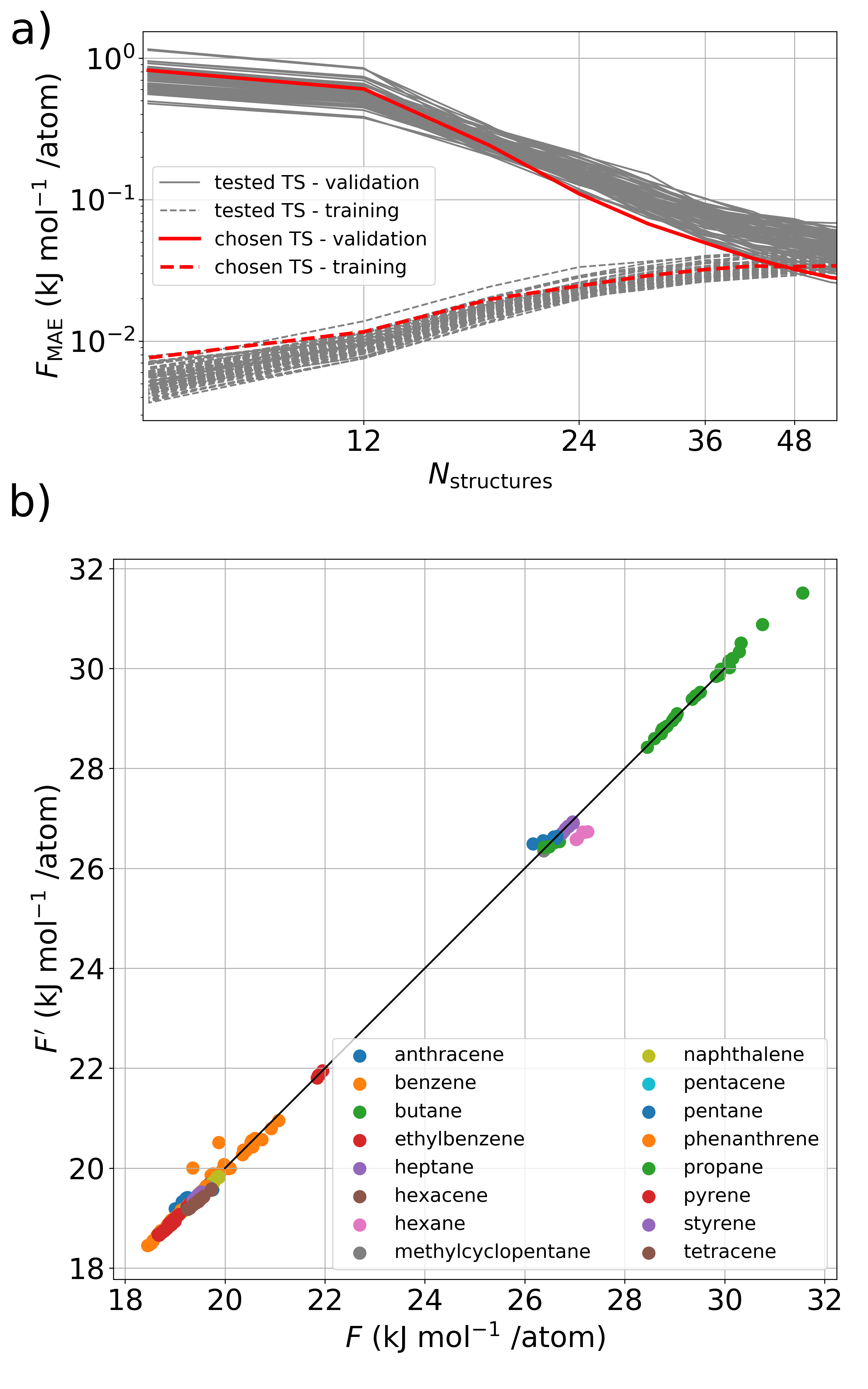}
    \caption{\label{fig:model}
    a) Learning curves obtained for 100 randomly chosen learning sets out of total 444 in grey, and the chosen learning set with the highest recorded value of $I(N_{m})$ in red.
    Data was obtained using SOAP representation and the classical force field model with the free energy calculated at 300K.
    b) Correlation between predicted and calculated free energies at 300 K (classical force field, SOAP representations).
    Different crystal families are represented by different colors.
    }
\end{figure}

In the same spirit of maximizing accuracy and minimizing cost, with the objective of performing free energy predictions with \textit{ab initio} data, the aim was to select an efficient and reliable validation set, without using the entire ensemble.
Here the goal is to create such a subset that would represent well the entire set, so as to include, for example, a proportional number of outlier structures as found in the entire set.
A random selection of validation set would not fulfill this criterion due to the limited size of validation set used in this project.
Additionally, this task largely differs from selecting the training set, because it typically contains a greater relative number of outliers compared to the entire set.  
In order to optimally select the validation set, while preserving the density of outliers, a stratified approach was used.
Here each crystal structure $a$ is assigned a similarity index $S_a$, that compares a given crystal to entire set
\begin{eqnarray}
    S_{a}= 
            \sum_b^{N_{\Omega}}{R_{a\rightarrow b}}.
    \label{eq:sim_index}
\end{eqnarray}

The relatively high values of $S_a$ indicate a ``typical'' crystal and low values indicate ``outliers''.
Next, the entire set is sorted with respect to  $S_a$ and the validation set is chosen by selecting every $n$-th element of the sorted set, with $n=round(N_{\Gamma}/N_{\Theta})$ where $N_{\Gamma}$ and $N_{\Theta}$ are the target numbers of structures in the validation and training sets.
All sets sorted with respect to $S_a$ are presented in supplementary information Figure SI2.

Within the discussed framework and common to many ML models, the choice of method encoding the atomic environments to numerical representations has an impact on the resulting performance of the model.
In this project, three well-established general-use atomic environment representations \cite{langer2020representations} were selected and tested, namely: Smooth Overlap of Atomic Positions (SOAP) \cite{C6CP00415F} that uses spherical harmonics to locally expand atomic densities, Many Body Tensor Representation (MBTR) \cite{mbtr} that uses distributions of different structural motifs (like radial or angular distribution functions) and Atom-centered Symmetry Functions (ACSFs) \cite{acsf} that use two- and three-body functions detecting specific features.
The Python implementations of the mentioned representations found in the DScribe package\cite{dscribe} was used.

\subsection{\label{sec:ts_selections}Model implementation and validation}

For the purpose of this work we have chosen crystals composed of seventeen different hydrocarbons: pyrene, methylcyclopentane, styrene, naphthalene, benzene, tetracene, mesitylene, pentane, pentacene, hexane, ethylbenzene, propane, heptane, phenanthrene, butane, hexacene and anthracene. 
We have included most available polymporphs that could be obtained from the Cambridge Crystallographic Data Centre \cite{ccdc} (CCDC), leading to an ensemble of 74 structures.
We noted that polymporphs of very similar lattice constant in CCDC tend to be almost identical, with close to negligible differences in atomic positions, for example, the case of ANTCEN20 and ANTCEN22.
Finally, the sample domain was further expanded by introducing structures with perturbations of roughly 5\% in the lattice parameters, as this can lead to up to 16\% increase in unit cell volume - a typical volume expansion percentage for molecular crystals~\cite{Beran2016}.
The addition of crystal structures with strongly perturbed lattice parameters was found to be crucial for the later prediction of lattice expansion coefficients.
Finally, $N_{\Omega}=444$ crystal structures were considered in this project.

The building and testing of the framework was initially performed using a classical force-field potential (AIREBO, as detailed in Methods).
In the first steps of the model verification, the training and validation set selection criterion, based on the FPS method and maximization of $I(N_{m})$, was evaluated.
For this purpose, based on the classical force field data with prediction performed at 300 K and with SOAP atomic-environment representations, free energy mean absolute error $F_{\text{MAE}}$ was calculated
\begin{eqnarray}
    F_{\text{MAE}}= \frac{1}{N}\sum_i^{N}{|F_i-F'_i|},
    \label{eq:fmad}
\end{eqnarray}
where $N$ is the number of structures for which the prediction is performed.
The results are presented in Figure \ref{fig:model}a in the form of learning curves, with increasing size of the training set $N$ and with the validation set.
It is visible that the learning curve obtained for the chosen training set, so with the highest $I(N_{m})$, shows one of the lowest $F_{\text{MAE}}$ at the target training set size among all potential sets obtained using FPS method.

Next, the linear and monotonic correlations between benchmark $F$ and predicted $F'$ values was assessed by calculating the Pearson and Spearman correlation coefficients.
For predictions performed at 300 K with the SOAP representation they were found to be 0.9996 and 0.9894, respectively.
A value so close to 1 for these coefficients indicate a good performance of the developed framework.
Furthermore, due to the low cost of the lattice dynamics calculations performed using classical force field, the $F_{\text{MAE}}$ was inspected for the entire set (300 K with the SOAP representation) and it was found to be 0.042 kJ/mol/atom.
Additionally, the $F_{\text{MAE}}$ = 0.218 kJ/mol/atom was obtained for 10\% of the crystals with the poorest prediction  and 0.023 kJ/mol/atom for the remaining 90\% of samples.

Figure \ref{fig:model}b shows the predicted free energy values $F'$ compared with the benchmark data $F$ for the different crystal families.
The analysis gives an indication of the system-sensitive performance of the framework, revealing that crystals of pentane, pentacene, tetracene and hexane are characterised by the poorest averaged prediction accuracy, with the $F_{\text{MAE}}$ around $2$ kJ/mol/molecule, reaching a possible free energy difference between different polymporphs \cite{C5CE00045A, C6CP05447A}.
Additionally, the predictions performed for crystal structures with strongly perturbed lattice parameters were noticeably poorer, even if the training set contained parental crystal structures.
Nevertheless, the prediction accuracy overall is very high, especially considering the diversity of hydrocarbons represented.

\begin{figure}[ht!]
    \includegraphics[width=0.45\textwidth]{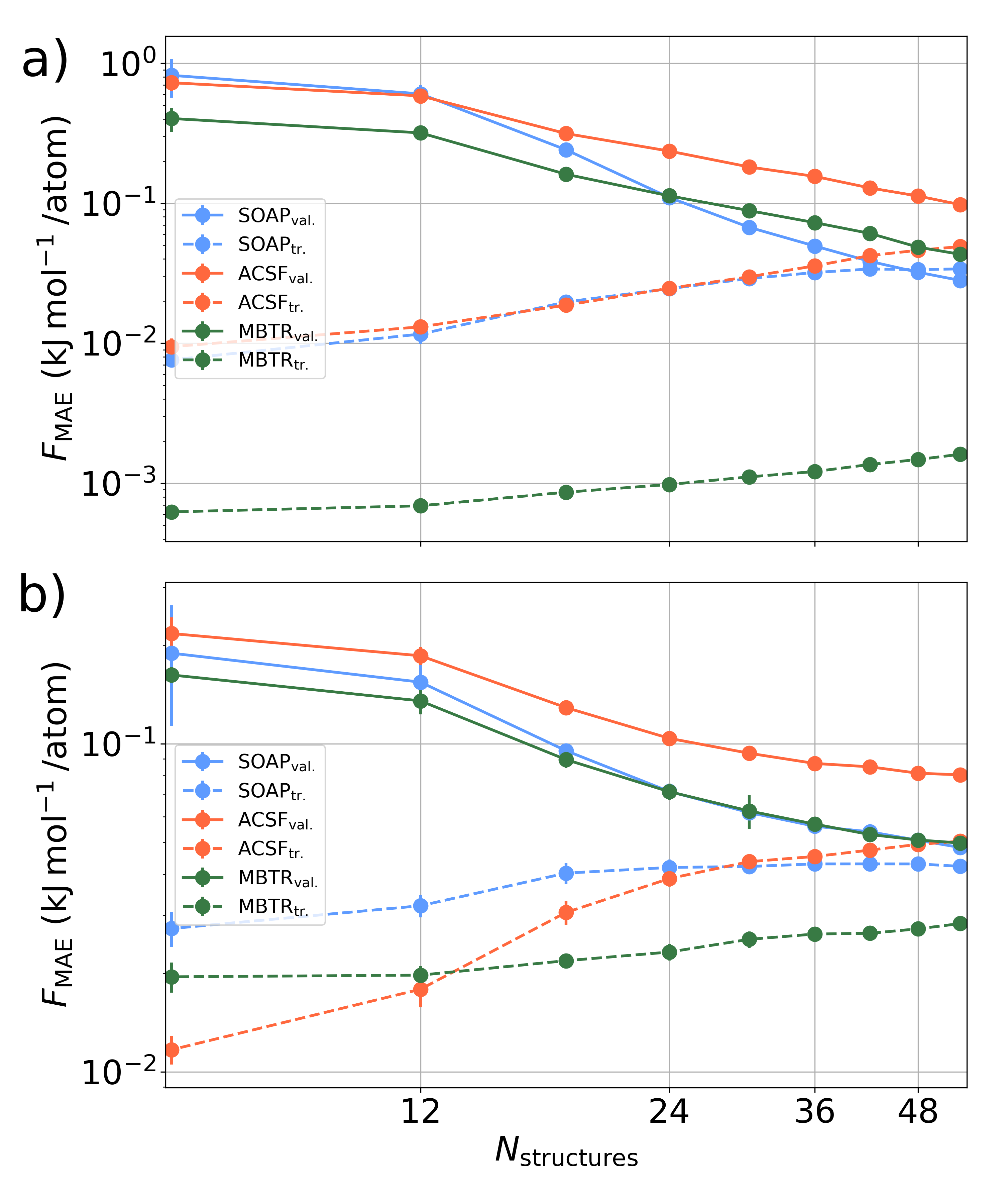}
    \caption{\label{fig:fmad}
    Learning curves of $F_{\text{MAE}}$ (300 K) calculated for the training (dashed line) and the validation set (solid line) obtained with SOAP, MBTR and ACSF representations.
    Results are presented for GPR models obtained based on: a) the empirical AIREBO force field and b) density-functional theory (PBE functional with pairwise van der Waals corrections) data, and are presented as a function of the number of crystal structures in the training set.
    Error bars are equal to the standard deviation of $F_{\text{MAE}}$ of training sets with different structures.
    }
\end{figure}

Figure \ref{fig:fmad}a shows the learning curves by monitoring $F_{\text{MAE}}$ at 300 K with increasing training set size and a constant validation set.
Additionally, the impact of the atomic environment representation on the efficiency of the method was investigated.
The learning is well-behaved for all representations, as expected for properly parameterized machine learning models.
The results obtained with the SOAP representation, with a 6 \AA~cutoff and 1 \AA~for the standard deviation of the Gaussian functions used to expand the atomic density, are characterised by the lowest $F_{\text{MAE}}$, showing that it is the best representation within the investigated set.
Finally, the accuracy of the predictions are noticeably affected by the temperature at which the free energies are required, going from 0.019 kJ/mol/atom at 300 K and 0.015 kJ/mol/atom at 200 K to as low as 0.002 kJ/mol at 0 K.

\subsection{Transferability of the prediction model}

\begin{figure}[htbp!]
    \includegraphics[width=0.45\textwidth]{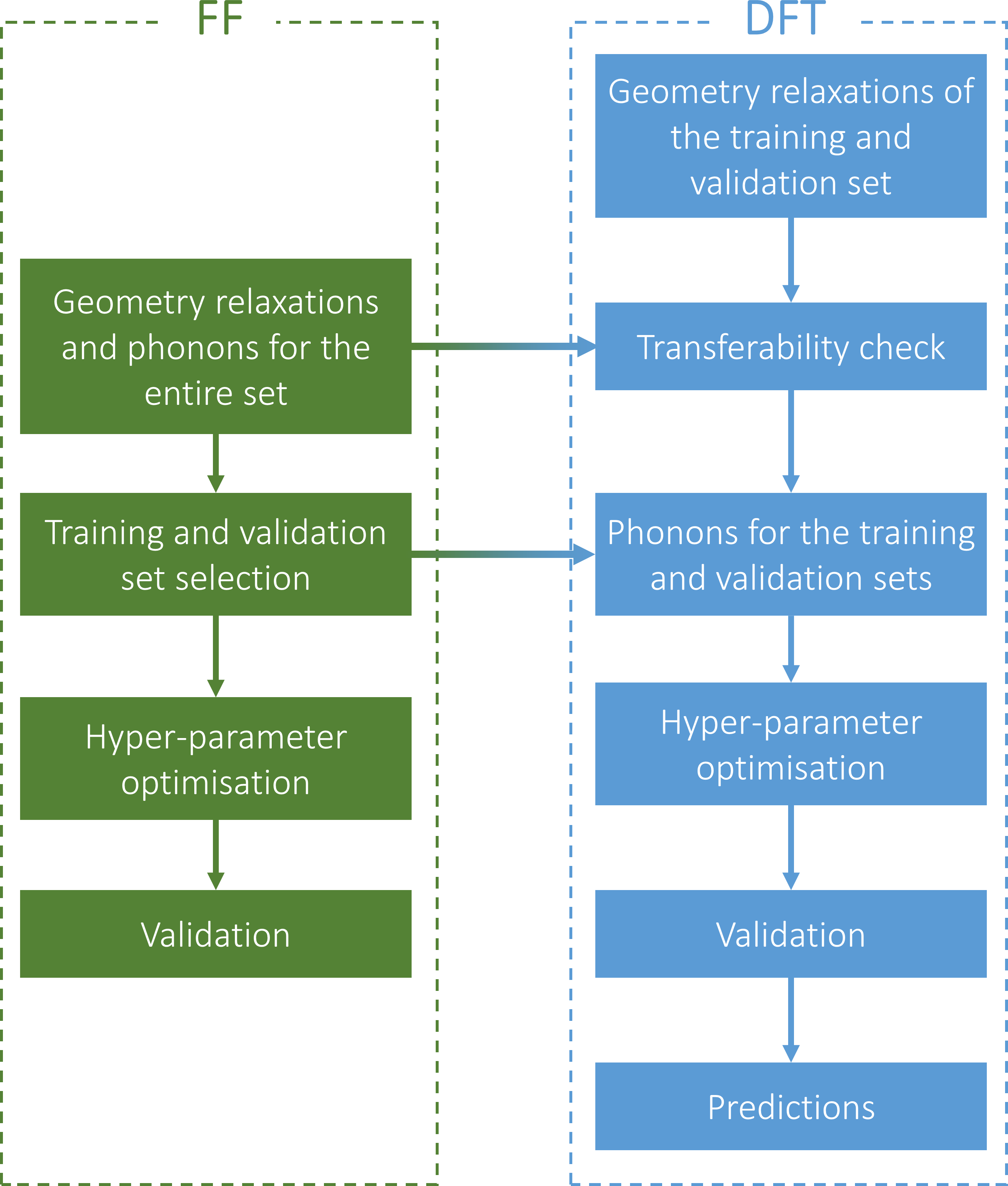}
    \caption{\label{fig:flow}
    Flowchart of the developed framework.
    }
\end{figure}

Once the framework was built and proven to deliver a satisfactory prediction of harmonic free energies based on data coming from an empirical potential, the transferability of the model when using DFT data was investigated.
For that, the PBE exchange correlation functional with pairwise van der Waals interactions was employed, as detailed in Methods.
As a test, the similarity between relaxed structures obtained with the empirical potential and DFT was assessed by analysing the root mean square deviation (RMSD) of the atomic positions averaged over entire set.
RMSD for carbon and hydrogen were $0.16$ \text{\AA} and $0.20$ \text{\AA}, respectively.
Importantly, differences in the SOAP representation were also investigated by calculating the root mean square error normalized by the standard deviation $\epsilon_X$, defined as
\begin{eqnarray}
\epsilon_{X} = 100 \times
    \sqrt{
        \frac{
                 \sum_d^D \frac{1}{N}\sum_{i}^{N} \left(q_{d,i}^{FF}-q_{d,i}^{DFT} \right)^{2}
            }{
                 \sum_d^D \frac{1}{N-1}\sum_{j}^{N} \left(\bar{q}_{d}^{DFT}-q_{d,i}^{DFT} \right)^{2}
            }
        }
\end{eqnarray}
where $D$ is the number of features of the representation and $N$  is the number of atoms for which the $\epsilon_X$ was calculated.
Obtained values for both carbon and hydrogen are $\epsilon_{C}$=1.09 and $\epsilon_{H}$=1.15 respectively.
Those results show that the overall structural features are in good agreement in these two potential energy surfaces.
As a consequence of this structural similarity between two data sets, the training and validation sets obtained with the empirical potential, as explained in the previous section, can be automatically used in DFT.
As a cross-check, the same training and validation set optimization procedure were independently applied on the optimized DFT structures, indeed obtaining the same results.
This proved that the experience gathered from the first phase of the project, where only classical data was used, is fully transferable to the current stage, where we employ more accurate \textit{ab initio} data.
As a result, the more expensive \textit{ab initio} lattice dynamics calculations were only performed for crystal structures included in the training and validation sets, greatly reducing the computational cost of the model generation.

Finally, the hyperparameters of the GPR model were re-optimised and were used to calculate learning curves for training and validation sets shown in Figure \ref{fig:fmad}b, with the SOAP, MBTR and ACSF representations.
All representations presented a good performance, with MBTR and SOAP yielding very similar learning curves. 
The obtained $F_{\text{MAE}}$ for the SOAP representation at full training set was found to be 0.038 kJ/mol/atom.
Interestingly, a fairly good prediction performance can be obtained with as little as 20 crystal structures, resulting in $F_{\text{MAE}}$=0.07 kJ/mol/atom.
Such small training sets typically do not contain all different molecular components of the crystals that are present in the entire set, but can still describe it well.
The remainder of this manuscript will focus on results obtained based on the DFT data with the SOAP representation, exclusively. 

The proposed framework is summarised in the flowchart in Figure \ref{fig:flow}.
In addition, as it is shown in the SI, the possibility of this model trained only on hydrocarbons to extrapolate to systems containing carbon, hydrogen and nitrogen atoms was investigated.
Although the prediction accuracy decreases as the concentration of nitrogen atoms in the samples increases, the model is not completely invalid.
It shows that with a small addition of structures to the training set, or building representations for new atoms that combine characteristics of the atoms that were previously trained \cite{Grisafi2019, van_der_Giessen_2020} this framework could be easily extended to other systems.

\subsection{Relative free energies of molecular crystals: stability ranking}

\begin{figure}[ht!]
    \centering
    \includegraphics[width=0.45\textwidth]{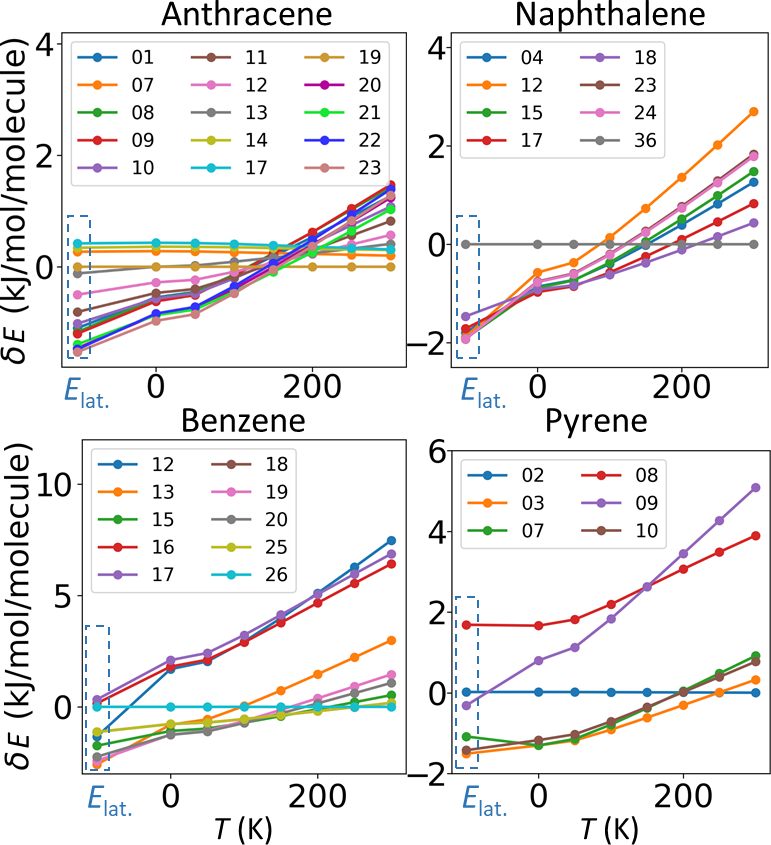}
    \caption{\label{fig:polimorphs}
    Lattice energy $E_{lat.}$ and predicted free energy differences between variants of four molecular crystal families: anthracene, naphthalene, benzene and pyrene.
    Full data, also for a larger number of crystals and polymporphs, is available in the Table S2 in the SI.
    Relative energies are calculated separately at each temperature, always with respect to the lowest free energy structure at 300K.
    The numbers on the labels represent the identifiers of the crystals following the convention used in CCDC \cite{ccdc}.
}
\end{figure}

The GPR model was employed to create a stability ranking of several families of hydrocarbon molecular crystals.
Sixteen crystal families were considered, encompassing 
38 polymorphs and 36 variants corresponding to different thermodynamical conditions with lattice parameters as they are given by the CCDC \cite{ccdc}.
Additionally, 370 crystal structures with randomly distorted lattice parameters derived from the initial 74 were included.
Figure \ref{fig:polimorphs} shows the lattice energy and the free energy obtained at various temperatures, presented as relative values to the crystal structure  characterised by the lowest free energy at $300$ K (full data is found in Table S2, in the SI).
The identifiers of all crystal structures follow the convention used in CCDC \cite{ccdc}.
For many crystal families, the structure with the lowest lattice energy is not the same as the one with the lowest free energy\, especially at the room temperature.
A clear example is the pyrene crystal and its three polymorphs: Form I  is represented by PYRENE02 and PYRENE03 (structures measured at 423 K and 113 K, respectively, and ambient pressure); Form II is represented by PYRENE07 and PYRENE10 (at 93 K and 90 K, ambient pressure); and Form III is represented by PYRENE08 and PYRENE09 (measured at at 0.3 GPa and 298 K, and at 0.5 Gpa and 298 K, respectively). Form I is measured to be more stable than form III at all temperatures up to and beyond 430 K, at ambient pressure.
Here, it is shown that the energy ranking formed based on lattice energy exclusively would place the high-pressure form III PYRENE09 (form III) structure very close to PYRENE02 (form I).
An inclusion of zero-point-energy and vibrational contributions already at low temperatures irrevocably destabilizes form III.

A similar example is the benzene crystal.
Here, structures of the ambient-pressure form I, represented by, for example, BENZEN15, BENZEN19 or BENZEN26, are characterised by overall lower free energy comparing to the high-pressure form II structures, like BENZEN16 and BENZEN17.
Interestingly, for this crystal family,  the lattice energy can provides a satisfactory relative stability ranking.
However, the need for including the vibrational contributions becomes visible once a high and ambient pressure variants of one polymorph are compared, e.g. BENZEN13 and BENZEN26.
It is visible in Figure \ref{fig:polimorphs} that if considering only lattice energies, BENZEN13 shows the lowest energy compared to other crystal variants, with lattice energy lower than that of BENZEN26 by 2.58 (kJ/mol/molecule).
However, the free energy prediction shows that at $300$ K, the BENZEN26 structure becomes the most stable out of all those investigated, and its relative free energy with respect to BENZEN13 is now lower by 2.92 (kJ/mol/molecule), effectively swapping places in the relative ranking stability with BENZEN26.
For this case, and to test the predictions of the model in practice, the free energies for both BENZEN13 and BENZEN26 structures were additionally calculated with DFT.
These calculations showed that BENZEN13 is characterised by a free energy that is 3.43 (kJ/mol) higher than that of BENZEN26 at $300$ K, confirming the results obtained with the GPR model.

The rearrangement of the relative stability ranking when room temperature free energy is taken into account is a very common trend among the investigated samples, and there are a number of cases, where even at $0$ K the zero point energy contribution is high enough to affect the relative stability ranking.
These observations are in good agreement with previous studies, where more direct methods were used \cite{C5CE00045A}.
In some cases, the prediction accuracy of this model is not sufficient to determine the relative stability of some structures.
Nevertheless, the model is accurate enough to point towards those few that are characterised by the lowest free energies.
Here, even only narrowing the pool of considered strucrures can effectively decrease the computational effort of phonon calculations required, if more accuracy is needed.

\subsection{Predicting lattice expansion.}

\begin{figure}[ht!]
    \includegraphics[width=0.45\textwidth]{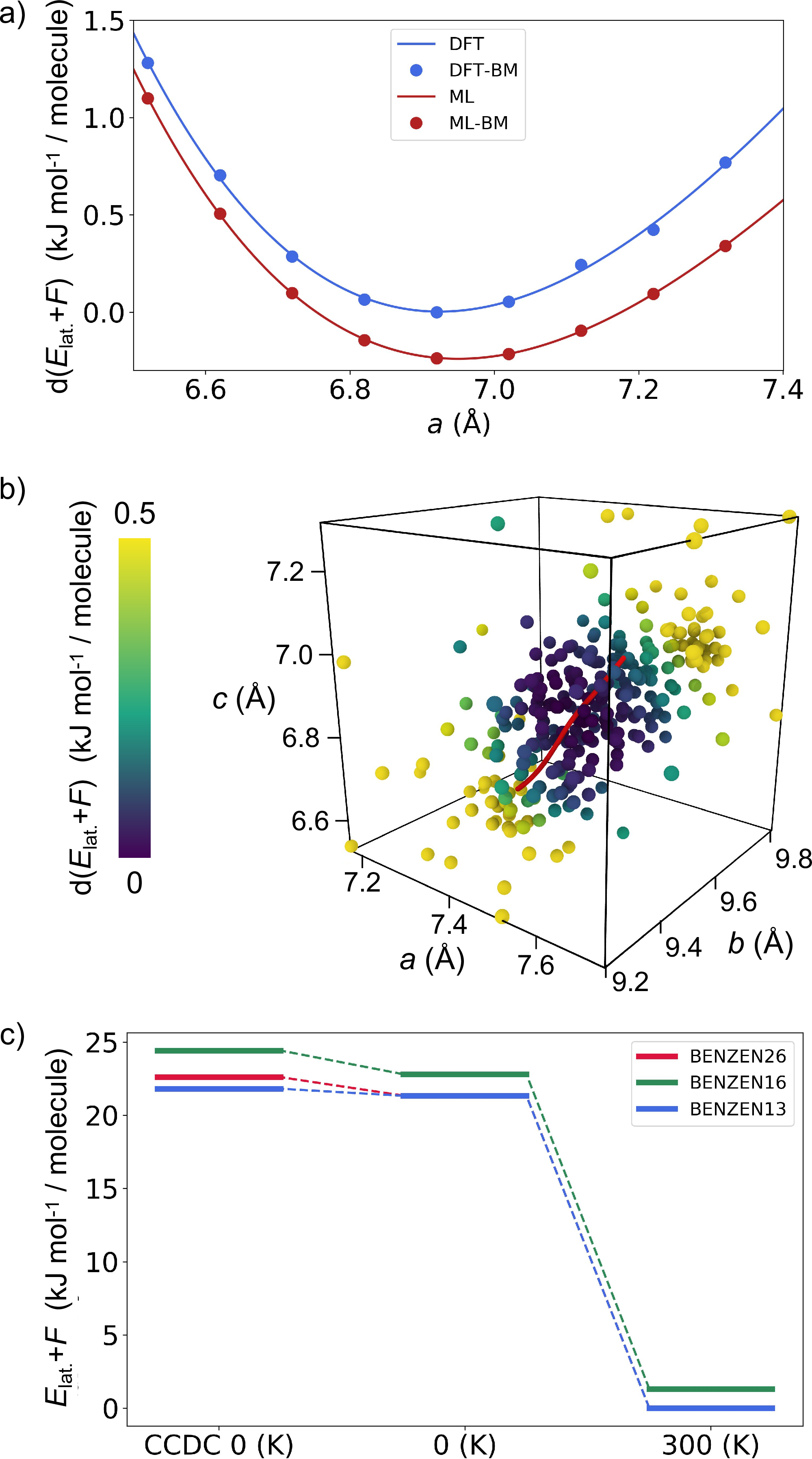}
    \caption{\label{fig:last}
    a) Predicted (red) and calculated (blue, DFT) free energies as a function of one lattice parameter of the BENZEN11 crystal at 200K. Solid lines correspond to a Birch–Murnaghan fit.
    b) 3D visualisation, in lattice parameter space ($a$, $b$, $c$),  of the free energy prediction including lattice expansion of the anthracene crystal at 200 K, with 300 different combinations of perturbed lattice parameters.
    The red line indicates the observed change in the lattice parameters when increasing the temperature from 0 to 300 K.
    c) Relative predicted free energies of benzene crystal structures when considering fixed lattice constants (taken from CCDC), and when considering lattice expansion at 0 K and at 300 K.}
\end{figure} 

Because one of the challenges in high throughput  computational screening of crystal structures is accounting for thermal lattice expansion, the application of the trained free energy model was explored in this context.
To illustrate the procedure, a simple case where only one of the lattice parameters is being perturbed was considered.
For this purpose the BENZEN11 crystal was chosen, with the lattice parameter $a$ being sampled within $6.52$ \text{\AA} and $7.32$ \text{\AA}.
Next, within the quasi-harmonic approximation, the free energy was calculated and predicted as a function of $a$.
Figure \ref{fig:last}a shows the comparison between the GPR model and DFT calculations for the free energy at $200$ K.
The optimal lattice parameter $a$ is determined by a Birch–Murnaghan \cite{Murnaghan244} fit.
While there are small differences between the DFT and the GPR curves, mostly consisting of a shift in energy, the resulting optimal lattice parameter $a$ is very similar in both cases, and equal to $6.92$ \text{\AA} and $6.95$ \text{\AA} respectively.
This simple and fairly artificial example illustrates that the prediction accuracy of this framework is sufficient to be employed in the context of the lattice expansion/contraction prediction.

A more challenging task is the prediction of anisotropic lattice changes.
Direct calculations of anisotropic lattice expansion requires lattice dynamic evaluations for, typically, hundreds of structures of the same crystal polymorph, making it a very costly calculation for a high-throughput setting.
Although harmonic and quasi-harmonic models \cite{PhysRevMaterials.3.053605} as well as an approach based on the assumption of the linear relation between free energy and volume have been proposed to overcome this cost \cite{C6CP02261H}, with this framework these lattice changes can be estimated without relying on any \textit{ansatz} for the dependence of the free energy on the lattice parameters.
It is worth noticing that even if the free energy predictions at various temperatures requires training the ML model multiple times, it happens with minimal computational overhead once appropriate lattice vibrations have been computed.
Four molecular crystals were picked, namely $P2_1/a$ anthracene (ANTCEN), $Pbca$ benzene (BENZEN), $P1$ pentacene (PENCEN01) and $Pbcn$ styrene (ZZZTKA01) and hundreds of ionic relaxations with the $a$, $b$ and $c$ lattice parameters perturbed by around 5\% were performed.
Next, for each of those perturbed structures free energy prediction at a number of temperatures from $0$ K to $300$ K range was performed.

Figure \ref{fig:last}b shows a 3D visualisation of free energy predictions for over 300 different combinations of lattice parameters $a$, $b$ and $c$ of the ANTCEN crystal.
Even with such a high number of sampled lattice parameter combinations, the position of the free energy local minima might not overlap with the gathered data.
In this case, in order to find the minimum in this high-dimensional space, an active learning based on the GPR algorithm is employed.
Here, the GPR is used as a multi-dimensional, non-linear regressor, as implemented in the scikit-learn \cite{scikit-learn} package.
In detail, the following bootstrap procedure is used:
\begin{enumerate}
  \item Identifying the position of the data point with the lowest free energy value according to the GPR 3D interpolation.
  \item For the chosen set of $(a,b,c)$ lattice parameters perform an ionic relaxation and predict the free energy with the trained model.
  \item If the predicted free energy of the $(a,b,c)$ sample varies from the free energy obtained by the 3D GPR regression, a new 3D GPR regression is performed, now explicitly including sample $(a,b,c)$, then go back to step 1.
  \item If the predicted free energy of the $(a,b,c)$ sample is sufficiently close to the one of the 3D GPR regression (within $\pm 0.1$\%), then the scheme is stopped and the optimal lattice parameters are considered to be found. 
\end{enumerate}

We found that typically only around 3 additional relaxations and free energy predictions (per temperature) are necessary to achieve sufficient convergence of the lattice parameters.
By employing this procedure to predict the anisotropic lattice changes the lattice-parameter change is calculated, as well as the full volume change of the selected crystals, as shown in Figure S5.

The results obtained can be compared to experimental values where data is available.
For anthracene the experimentally measured volume change is $V_{290K}^{exp.}/V_{90K}^{exp.}=1.024$ \cite{Masona04170} and we obtained $V_{290K}^{ML}/V_{90K}^{ML}=1.034$. For pentacene, the comparison is $V_{295K}^{exp.}/V_{90K}^{exp.}=1.037$ \cite{Mattheus2001} and $V_{295K}^{ML}/V_{90K}^{ML}=1.031$; for benzene $V_{270K}^{exp.}/V_{78K}^{exp.}=1.089$ \cite{benzen_1} and $V_{270K}^{ML}/V_{78K}^{ML}=1.068$; for styrene  $V_{120K}^{exp.}/V_{83K}^{exp.}=1.017$ \cite{Bondcf6127, Yasudaob6089} and $V_{120K}^{ML}/V_{83K}^{ML}=1.009$.
The predictions are quite close to experimental data and overall a high degree of anisotropy is observed.
Moreover, a deviation from a linear behavior of the free energy change with respect to volume is observed, as shown in Figure S6.

This framework can thus be used to create the relative stability ranking including the thermal expansion effect on the free energy.
Here, one example of how this can impact the relative stability and crystal form of these systems is presented. 
For this purpose, BENZEN13 and BENZENB26 (high and low pressure variants of the $P 2_1/b 2_1/c 2_1/a$ benzene I polymorph \cite{Budzianowski:av5045}) are selected, as well as BENZEN16 (a high pressure $P2_1/c$ benzene II polymorph \cite{Katrusiak2010}).
The initial lattice constants were taken from the CCDC.
As shown in Figure \ref{fig:last}c, by simply searching for the free energy minimum at 0 K using the procedure described above, BENZEN13 and BENZEN26 were found to end up being characterised by almost identical (predicted) free energies and lattice constants.
Further inspection indicated that indeed the BENZEN13 and BENZEN26 structures converged to the same structure, and the same behavior was found at all investigated temperatures.
Even if somewhat expected, given that they are high and low pressure phases within the same crystal group and in the absence of any applied pressure it is natural that they both adopt the low-pressure structure, the fact that this result came from the model alone, and that the free energy predictions were able to capture this transition, shows that the method is robust.
The BENZEN16 structure is stabilized by 1 kJ/mol/molecule upon increasing the temperature from 0 K to 300 K, as shown in Figure \ref{fig:last}c.
This stabilization is accompanied by an appreciable lattice expansion with a volume increase of around 6\% from 0 to 300 K.

\section{Conclusions}
In summary, we proposed a framework provides a machine learning model with first principles accuracy for the harmonic Helmholtz free energies of molecular crystals, that is suitable for high-throughput studies. 
In addition, it was shown that the training and validation set of the model can be optimised using a cheaper empirical potential, and then transferred to first-principles calculations, thus substantially decreasing the cost of training, without sacrificing accuracy. 

The model was tested to predict the relative energetic stability ranking of several diverse hydrocarbon polymorphs and distorted crystal structures derived from them and the changes on this ranking with increasing temperature was studied.
We observed that in most cases, omitting thermal effects and instead using only the lattice energy, leads to misleading results.
Furthermore, it was shown that the model can be efficiently employed to calculate the anisotropic lattice expansion -- a task rarely approached due to its complexity and high computational demand when performed at the \textit{ab initio} level.
Unsurprisingly, taking the anisotropic lattice expansion into account leads to further changes in the stability ranking.
Naturally, the same framework could be used to predict other quantities derived from vibrational properties, like the vibrational heat capacity.

The strengths of this framework are its low computational cost, reliability and accuracy.
However, because the model is trained to directly predict free energies, one still has to deal with the computational cost of obtaining optimized structures, which here we obtained from first-principles geometry optimizations.
Nonetheless, fitting a machine-learned interatomic potential is becoming more streamlined \cite{McDonagh2019}, even though these potentials rarely target the accurate description of vibrational properties due to the added complexity of including them in the learning procedure.
The presented framework, on the other hand, can be easily combined with any potential that can predict structures in a reasonable manner and has the potential to be more accurate.

Extending this framework beyond hydrocarbon-based crystals could be  straightforward, albeit perhaps requiring different training data.
We have already observed that the framework is capable of predicting DFT free energies from FF-relaxed structures with promising accuracy (see Supplementary Information).
Finally, targeting fully anharmonic free energies with \textit{ab initio} accuracy is still a daunting task that can, nevertheless, profit from the knowledge gained in this study.

\section{Methods}
Geometry optimisation calculations with empirical potentials were performed using LAMMPS \cite{lammps} together with AIREBO \cite{airebo} interatomic potentials.
The conjugate gradient minimization algorithm was used with dummy parameters to ensure full convergence, namely $10^{-25} (1)$ and $4 \times 10^{-25}$ kJ/mol/\AA~ for energy and forces respectively and with $5 \cdot 10^{4}$ maximum iterations of the minimizer.
Phonon calculations with the empirical potentials were performed using the i-PI \cite{ipi} code, considering $2 \times 2 \times 2$ repetitions of the primitive cell. The phonons were calculated by  finite differences with a  0.005 \text{\AA} displacement in all Cartesian directions.

All \textit{ab initio} simulations were performed using the FHI-aims package \cite{fhiaims}.
For this purpose, we employed \textit{light} settings for all atomic species, together with the Perdew-Burke-Ernzerhof exchange-correlation functional \cite{pbe} and many-body dispersion corrections \cite{mbd}.
We have used $5 \times 5 \times 5$ k-point sampling of the Brillouin zone. 
A self-consistency convergence criterion of $10^{-5}$ eV/\AA~ was imposed on the forces, which ensured that energies were converged to $10^{-7}$ eV or below.
The relaxation was performed using the trust radius version of the Broyden-Fletcher-Goldfarb-Shanno \cite{Pfrommer1997, bfgs} optimization algorithm with the maximum residual force component threshold equal to $10^{-4}$ eV/\AA.
Lattice dynamics calculations were performed through finite differences using Phonopy \cite{phonopy}. The atomic displacements were of $0.002$ \AA\, in all Cartesian directions.
The size of the supercell was individually chosen for the different molecular crystals, with the requirement that at least twice the distance between molecular centers of mass of adjacent molecules was comprised by the vector lengths in each direction.

The framework for the GPR model was developed using Python3 and Fortran95 languages.
A preliminary version of the core functionalities is available in \texttt{https://github.com/sabia-group/fep.git}.
The SOAP, MBTR and ACSF representations were calculated using the DScribe \cite{dscribe} package.

\section*{Data Availability}
All data necessary to replicate and interpret the free energy prediction framework discussed in this article can be accessed in the NOMAD repository with identifier \texttt{https://dx.doi.org/10.17172/NOMAD/2020.09.16-1}.

\section*{Acknowledgements}
We acknowledge useful discussions with T. Bereau, M. Langer, L. Ghiringhelli and M. Ceriotti. We thank M. Rupp and M. Langer for a critical read of the manuscript draft.
This work has been financially supported by BiGmax, the Max Planck Society's Research Network on Big-Data-Driven Materials-Science.

\section*{Author Contributions}
M.K. was responsible for designing, developing and programming the GPR framework, and performing all necessary simulations.
M.R. was responsible for the project planning, design and supervision. Both authors wrote the manuscript and analyzed the data.
\bibliography{ref}
\end{document}


\title{Efficient Gaussian Process Regression for prediction of molecular crystals harmonic free energies - supplementary information}

\author{Marcin Krynski$^{1,2}$*, Mariana Rossi$^{1,3}$}

\maketitle

1. Fritz Haber Institute of the Max Planck Society, Berlin, Germany

2. Warsaw University of Technology, Faculty of Physics, Warsaw, Poland

3. MPI for the Structure and Dynamics of Matter, Hamburg, Germany

* marcin.krynski@pw.edu.pl

\section{GPR Model: Parameters and Tests}

\begin{figure}[htbp]
    \centering
    \includegraphics[scale=0.45]{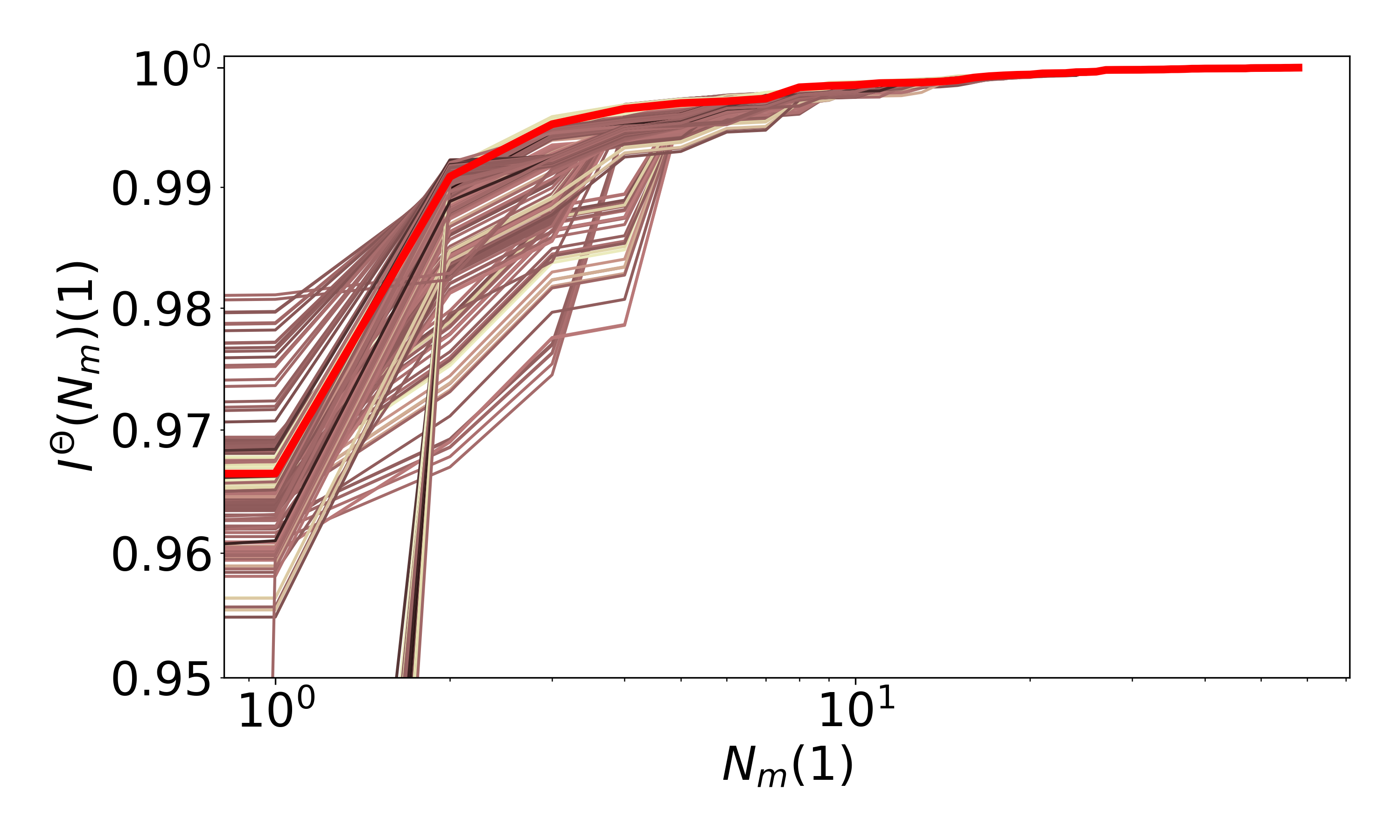}
    \caption{\label{fig:ls_coverage}
    Convergence of the $I^{\Theta}(N_{m})$ calculated for all potential training set candidates.
    Colours from yellow to brown marks sets of low to high $I^{\Theta}(N_{m})$ convergence.
    The red colour is marking the selected training set.
    }
\end{figure}

\begin{table}[htbp]
    \caption{\label{tab:descriptors}
    Results of the GPR hyper-parameters optimisation together with the mean absolute error of the harmonic free-energy calculated for all discussed descriptors based on both, \textit{ab initio} and classical model.
    All presented values were obtained based on the entire training set ($N_{m}=60$).
    }
    \centering
    \begin{tabular}{rcccc}
        \textrm{descriptor}&
        \textrm{$\sigma$}&
        \textrm{$l$}&
        \textrm{$\sigma_{\epsilon}$}&
        \textrm{$F_{MAE}^{N_m=60} (kJ/mol/atom)$}\\
        SOAP$_{DFT}$ & 0.077  & 72   & 0.03 & 0.038\\
        MBTR$_{DFT}$ & 0.077  & 414  & 0.03 & 0.040\\
        ACSF$_{DFT}$ & 0.074  & 18   & 0.03 & 0.063\\
        SOAP$_{FF}$  & 0.097  & 22   & 0.03 & 0.020\\
        MBTR$_{FF}$  & 0.097  & 157  & 0.01 & 0.033\\
        ACSF$_{FF}$  & 0.096  & 42   & 0.04 & 0.072\\
    \end{tabular}
\end{table}

\begin{figure}[htbp]
    \centering
    \includegraphics[scale=0.45]{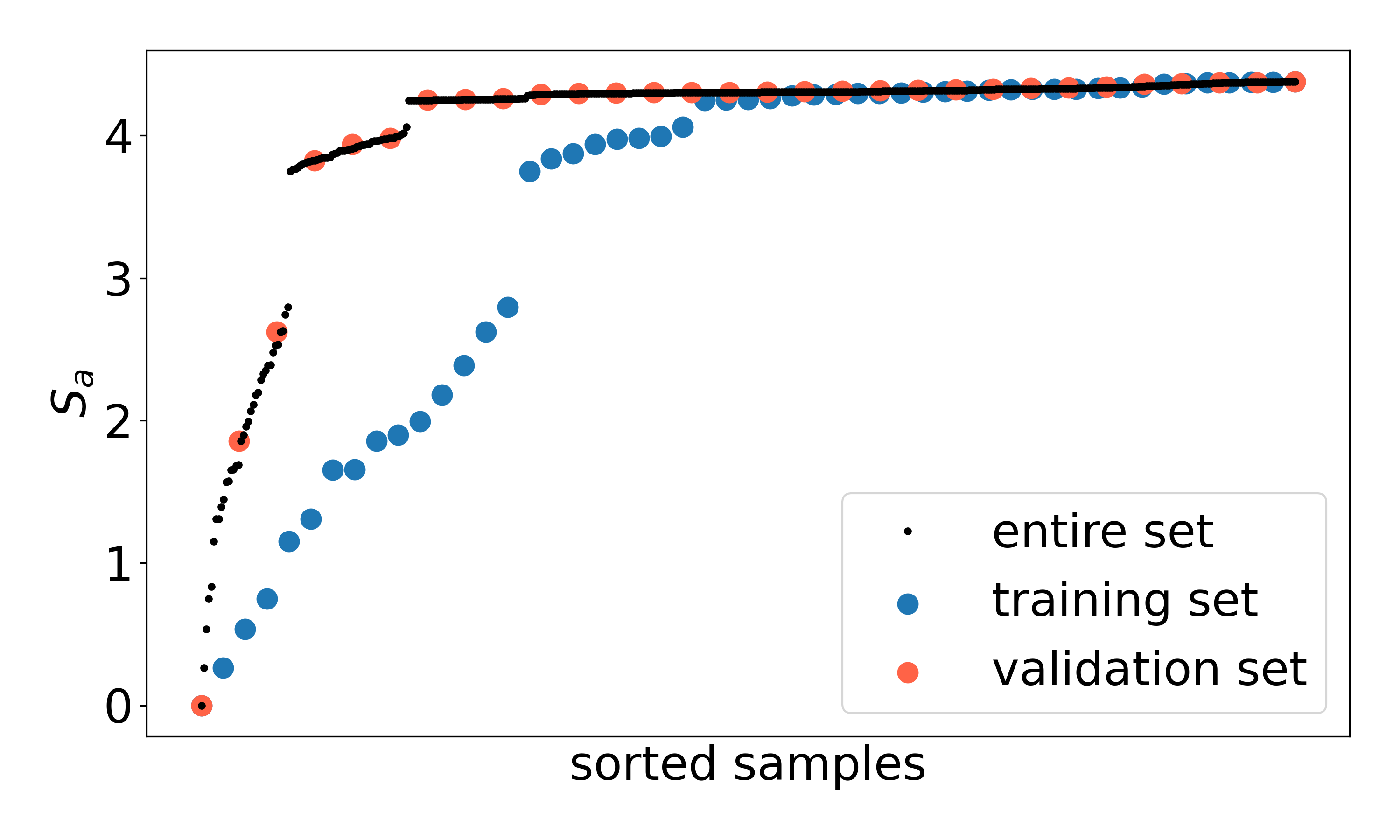}
    \caption{\label{fig:set_information}
        Sorted vector $S_a$ calculated for each of discussed sets: validation, training and the entire set.
        Each dot represent one molecular crystal.
        Presented in the Figure \ref{fig:set_information} sorted vector $S_a$ for the training set differs greatly from this of the entire set.
        This is caused by the necessity for the training set to cover proportionally more outliers than those present in the entire set.
    }
\end{figure}

\begin{figure}[htbp]
    \centering
    \includegraphics[scale=0.45]{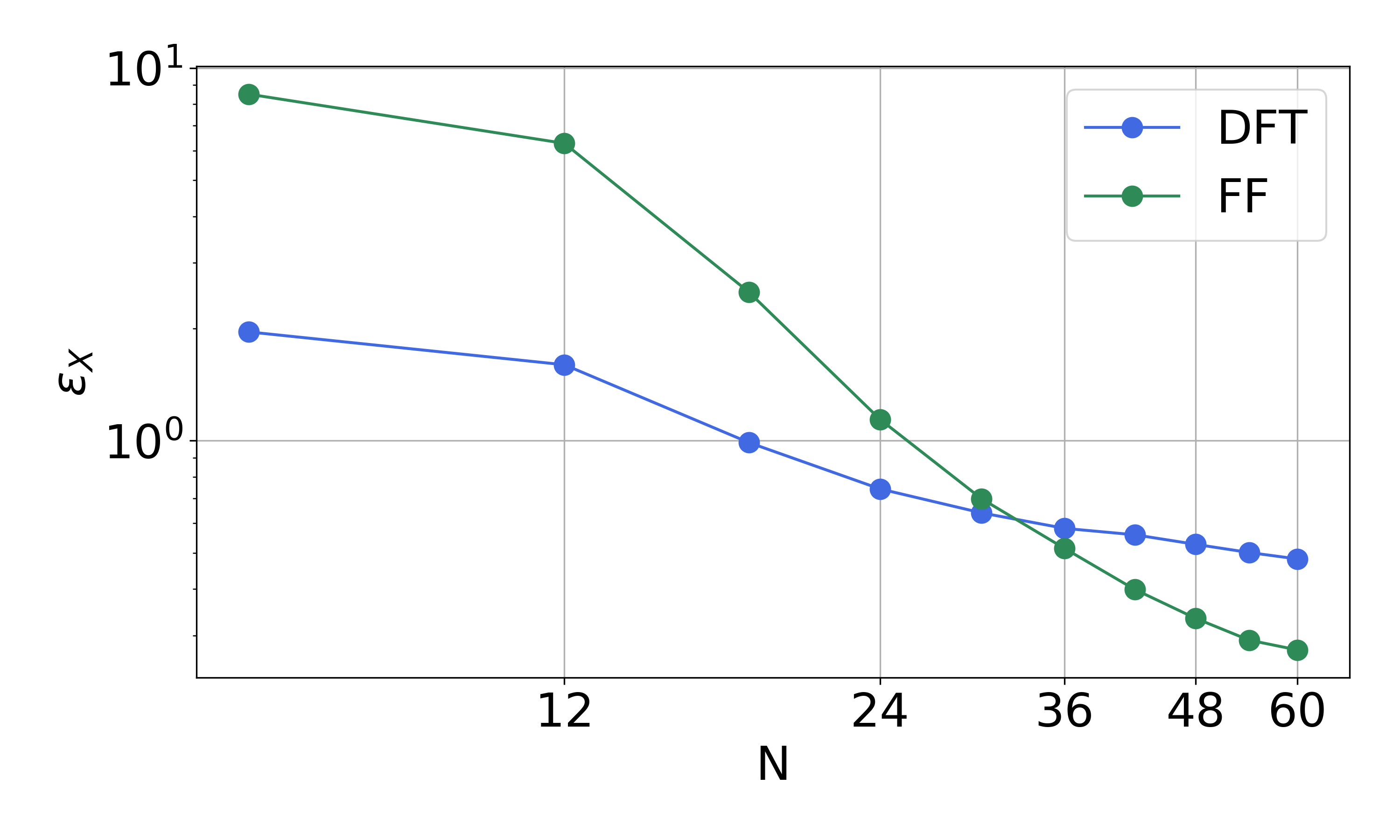}
    \caption{\label{fig:eps} Prediction error $\epsilon_{X} = 100 \times 
    \sqrt{
        \frac{
                \frac{1}{N_X}\sum_{i}^{N_X} \left(F'(X_{i})-F(X_{i}) \right)^{2}
            }{
                \frac{1}{N_X-1}\sum_{j}^{N_X} \left(\bar{F}-F(X_{i}) \right)^{2}
            }
        }$ (where $\bm{X}$ is the subset of $N_X$ crystal structures for which the prediction is performed) calculated for the validation and the training set for both, \textit{ab initio} and classical models at 300K.
        }
\end{figure}

\FloatBarrier
\section{Free Energy Predictions for Different Crystal Families}

\begin{center}
    \begin{longtable}[H]{cccccc}
        \caption{
        Lattice energy and free energy at 0K, 100K, 200K and 300K for chosen crystals structures of different families and polymorphs, calculated with respect to the structure showing the lowest free energy at 300K within a specific category.
        Structures identifiers are presented according to the CCDC data base format.
        All values are presented in $(kJ/mol/molecule)$.
        }\label{tab:ranking}\\

        \textbf{CCDC id.}&
        \textbf{$E_{lat}$}&
        \textbf{$F_{0K} $}&
        \textbf{$F_{100K}$}&
        \textbf{$F_{200K}$}&
        \textbf{$F_{300K}$}\\

		\endfirsthead
		\multicolumn{6}{c}
		{\tablename\ \thetable\ -- \textit{Continued from previous page}} \\

        \textbf{CCDC id.}&
        \textbf{$E_{lat}$}&
        \textbf{$F_{0K}$}&
        \textbf{$F_{100K}$}&
        \textbf{$F_{200K}$}&
        \textbf{$F_{300K}$}\\

		\endhead
		\multicolumn{6}{r}{\textit{Continued on next page}} \\
		\endfoot
		\endlastfoot
        \hline
        ANTCEN & 0.63 & 0.64 & 0.56 & 0.42 & 0.29\\
        ANTCEN01 & -1.10 & -0.55 & -0.15 & 0.53 & 1.27\\
        ANTCEN07 & 0.27 & 0.28 & 0.26 & 0.23 & 0.20\\
        ANTCEN08 & -1.17 & -0.57 & -0.14 & 0.62 & 1.43\\
        ANTCEN09 & -1.20 & -0.62 & -0.16 & 0.62 & 1.47\\
        ANTCEN10 & -1.02 & -0.57 & -0.21 & 0.41 & 1.08\\
        ANTCEN11 & -0.81 & -0.47 & -0.18 & 0.30 & 0.82\\
        ANTCEN12 & -0.50 & -0.28 & -0.09 & 0.23 & 0.57\\
        ANTCEN13 & -0.12 & -0.00 & 0.09 & 0.24 & 0.41\\
        ANTCEN14 & 0.34 & 0.36 & 0.35 & 0.32 & 0.30\\
        ANTCEN17 & 0.42 & 0.43 & 0.41 & 0.36 & 0.31\\
        ANTCEN19 & 0.00 & 0.00 & 0.00 & 0.00 & 0.00\\
        ANTCEN20 & -1.46 & -0.86 & -0.40 & 0.39 & 1.24\\
        ANTCEN21 & -1.39 & -0.87 & -0.45 & 0.26 & 1.03\\
        ANTCEN22 & -1.48 & -0.84 & -0.35 & 0.48 & 1.39\\
        ANTCEN23 & -1.53 & -0.97 & -0.48 & 0.37 & 1.28\\
        \hline
        BENZEN & 1.19 & 0.40 & 0.16 & 0.01 & 0.29\\
        BENZEN01 & 0.22 & 0.00 & 0.05 & 0.38 & 1.15\\
        BENZEN03 & 4.97 & 6.69 & 7.57 & 9.36 & 11.65\\
        BENZEN04 & 4.97 & 6.69 & 7.57 & 9.36 & 11.65\\
        BENZEN11 & 0.02 & 0.89 & 1.34 & 2.41 & 4.00\\
        BENZEN12 & 1.17 & 2.97 & 3.70 & 5.31 & 7.51\\
        BENZEN13 & -0.08 & 0.47 & 0.81 & 1.67 & 3.02\\
        BENZEN15 & 0.76 & 0.19 & 0.06 & 0.10 & 0.56\\
        BENZEN16 & 2.66 & 3.08 & 3.67 & 4.87 & 6.45\\
        BENZEN17 & 2.83 & 3.37 & 3.99 & 5.25 & 6.91\\
        BENZEN18 & 1.38 & 0.50 & 0.22 & 0.01 & 0.21\\
        BENZEN19 & 0.08 & 0.02 & 0.14 & 0.59 & 1.49\\
        BENZEN20 & 0.26 & 0.02 & 0.05 & 0.36 & 1.11\\
        BENZEN25 & 1.38 & 0.51 & 0.23 & 0.01 & 0.21\\
        BENZEN26 & 2.50 & 1.27 & 0.77 & 0.19 & 0.03\\
        \hline
        DUCKOB04 & 0.00 & 0.00 & 0.00 & 0.00 & 0.00\\
        DUCKOB05 & 2.53 & 9.30 & 10.87 & 13.97 & 17.31\\
        DUCKOB06 & 2.93 & 9.84 & 11.43 & 14.59 & 17.99\\
        DUCKOB07 & 7.38 & 15.62 & 17.38 & 20.99 & 24.93\\
        DUCKOB08 & 13.60 & 23.36 & 25.27 & 29.29 & 33.70\\
        DUCKOB09 & 19.90 & 31.08 & 33.09 & 37.39 & 42.14\\
        \hline
        HEPTAN01 & 0.00 & 0.00 & 0.00 & 0.00 & 0.00\\
        HEPTAN03 & -0.44 & 0.16 & 0.29 & 0.55 & 0.83\\
        \hline
        ZZZDKE01 & 0.00 & 0.00 & 0.00 & 0.00 & 0.00\\
        ZZZDKE02 & -0.58 & -0.31 & -0.14 & 0.17 & 0.51\\
        \hline
        NAPHTA04 & -1.80 & -0.86 & -0.40 & 0.39 & 1.26\\
        NAPHTA12 & -1.88 & -0.57 & 0.14 & 1.36 & 2.70\\
        NAPHTA15 & -1.91 & -0.88 & -0.37 & 0.51 & 1.48\\
        NAPHTA17 & -1.71 & -0.97 & -0.57 & 0.10 & 0.83\\
        NAPHTA18 & -1.46 & -0.92 & -0.62 & -0.11 & 0.44\\
        NAPHTA23 & -1.93 & -0.75 & -0.20 & 0.77 & 1.83\\
        NAPHTA24 & -1.92 & -0.77 & -0.22 & 0.74 & 1.79\\
        NAPHTA36 & 0.00 & 0.00 & 0.00 & 0.00 & 0.00\\
        \hline
        PENCEN & -0.11 & 0.19 & 0.41 & 0.84 & 1.35\\
        PENCEN01 & -2.14 & -1.10 & -0.55 & 0.50 & 1.71\\
        PENCEN05 & 3.33 & 3.65 & 3.27 & 2.70 & 2.17\\
        PENCEN10 & 0.00 & 0.00 & 0.00 & 0.00 & 0.00\\
        \hline
        JAYDUI & 0.00 & 0.00 & 0.00 & 0.00 & 0.00\\
        JAYDUI01 & 7.20 & 12.32 & 13.62 & 16.24 & 19.05\\
        \hline
        PYRENE & 0.02 & 0.02 & 0.02 & 0.01 & 0.00\\
        PYRENE01 & 0.00 & 0.00 & 0.00 & 0.00 & 0.00\\
        PYRENE02 & 0.02 & 0.02 & 0.02 & 0.01 & 0.00\\
        PYRENE03 & -1.51 & -1.30 & -0.91 & -0.30 & 0.32\\
        PYRENE07 & -1.08 & -1.30 & -0.79 & 0.05 & 0.92\\
        PYRENE08 & 1.69 & 1.66 & 2.19 & 3.07 & 3.89\\
        PYRENE09 & -0.31 & 0.80 & 1.83 & 3.45 & 5.08\\
        PYRENE10 & -1.42 & -1.17 & -0.71 & 0.02 & 0.77\\
        \hline
        ZZZTKA01 & -0.40 & -0.03 & 0.12 & 0.36 & 0.61\\
        ZZZTKA02 & 0.00 & 0.00 & 0.00 & 0.00 & 0.00\\
        \hline
        TETCEN & 0.00 & 0.00 & 0.00 & 0.00 & 0.00\\
        TETCEN01 & -1.14 & -0.69 & -0.43 & 0.02 & 0.51\\
        TETCEN03 & 2.76 & 2.81 & 2.33 & 1.55 & 0.77\\
	\end{longtable}
\end{center}

\newpage
\section{Free Energies Including Lattice Expansion}
\begin{figure}[htbp]
    \centering
    \includegraphics[scale=0.16]{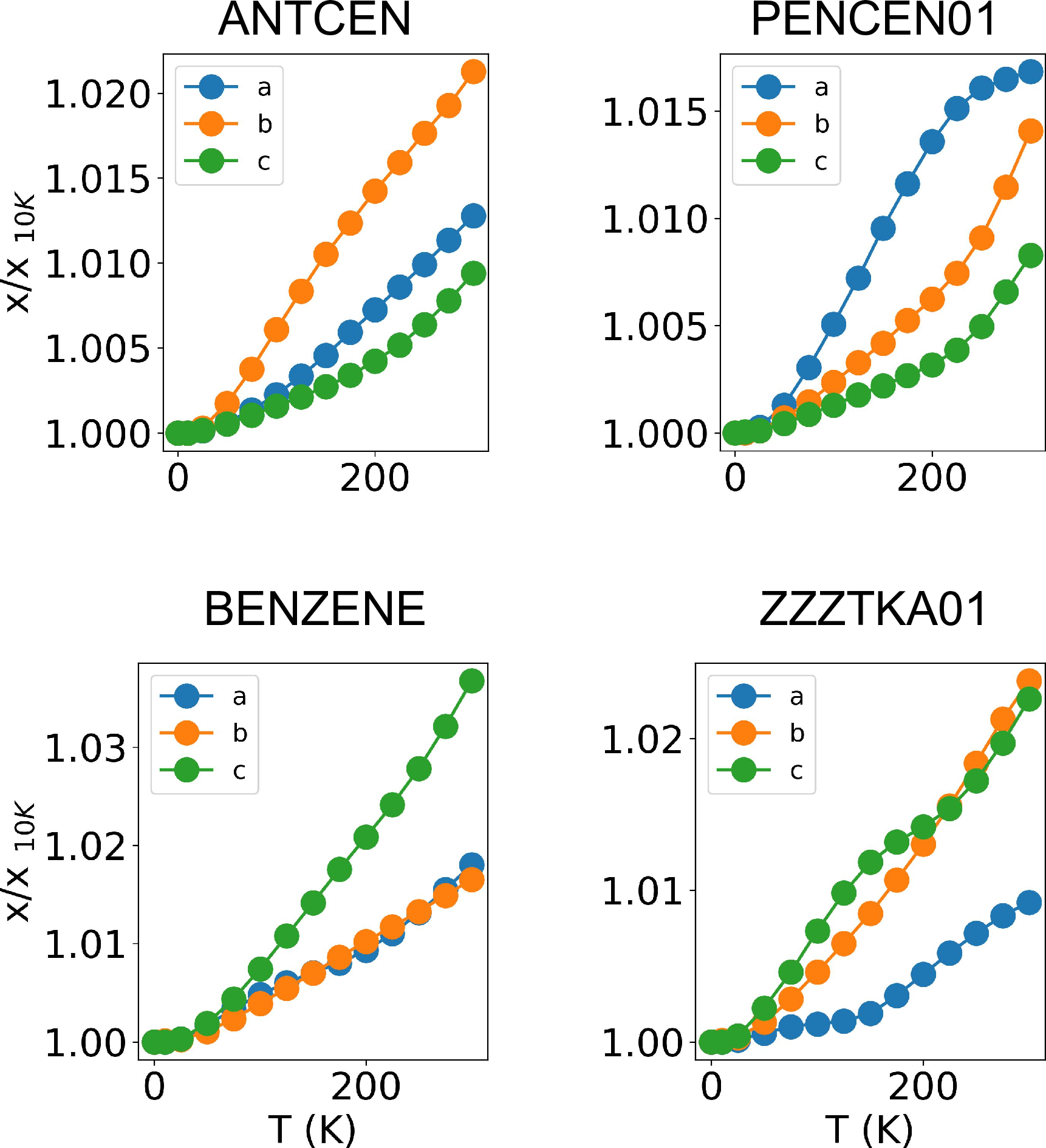}
    \caption{\label{fig:tle}
        Temperature evolution of the relative lattice parameters calculated for four investigated molecular crystals.
    }
\end{figure}

\begin{figure}[htbp]
    \centering
    \includegraphics[scale=0.5]{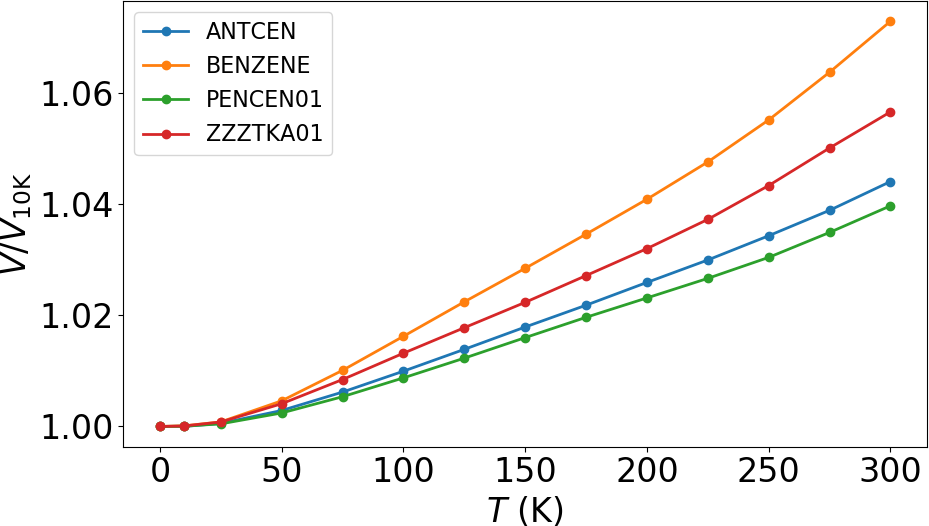}
    \caption{
        Temperature evolution of the relative volume calculated for four investigated molecular crystals.
    }
\end{figure}

\begin{figure}[htbp]
    \centering
    \includegraphics[scale=0.5]{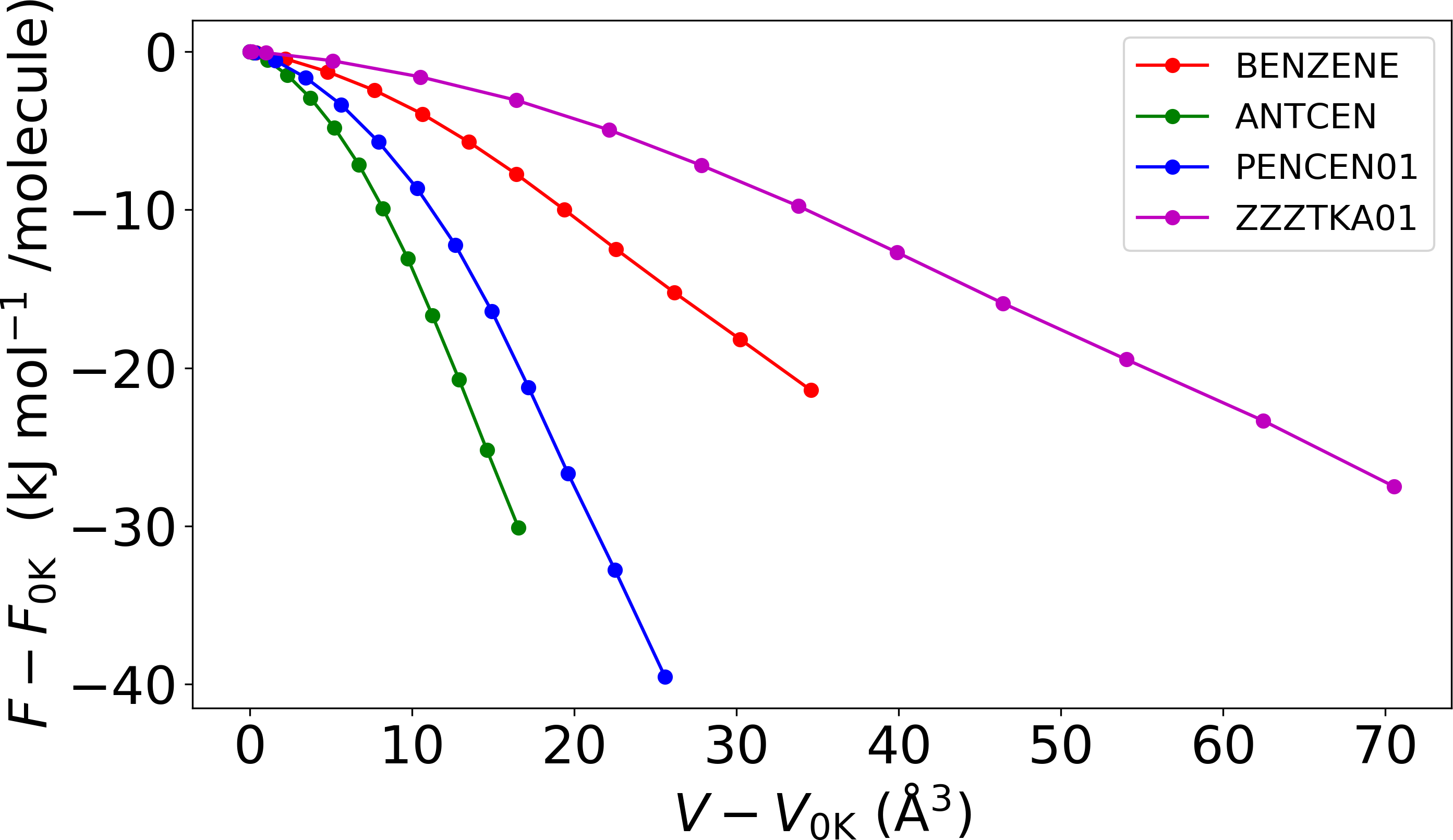}
    \caption{
        Relative free energy as a function of relative volume calculated within 0-300K range.
    }
\end{figure}

\begin{figure}[htbp]
    \centering
    \includegraphics[scale=0.5]{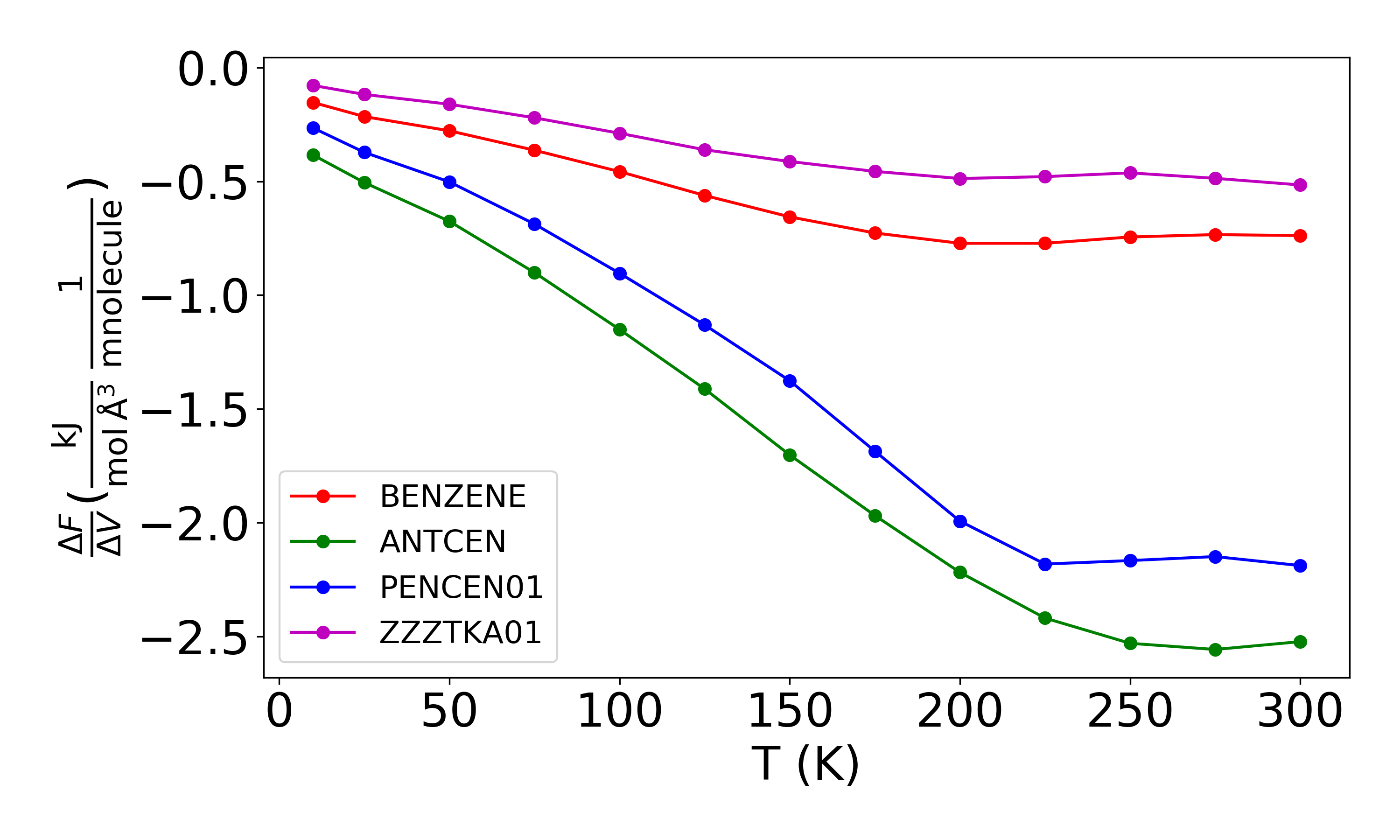}
    \caption{
        Temperature variation of the the thermal pressure 
        $P_{th}(T) = \frac{\Delta F(T)}{\Delta V(T)}$.
    }
\end{figure}

\FloatBarrier

\section{Predicting DFT free energies from force field structures}

\begin{figure}[htbp]
    \centering
    \includegraphics[scale=0.5]{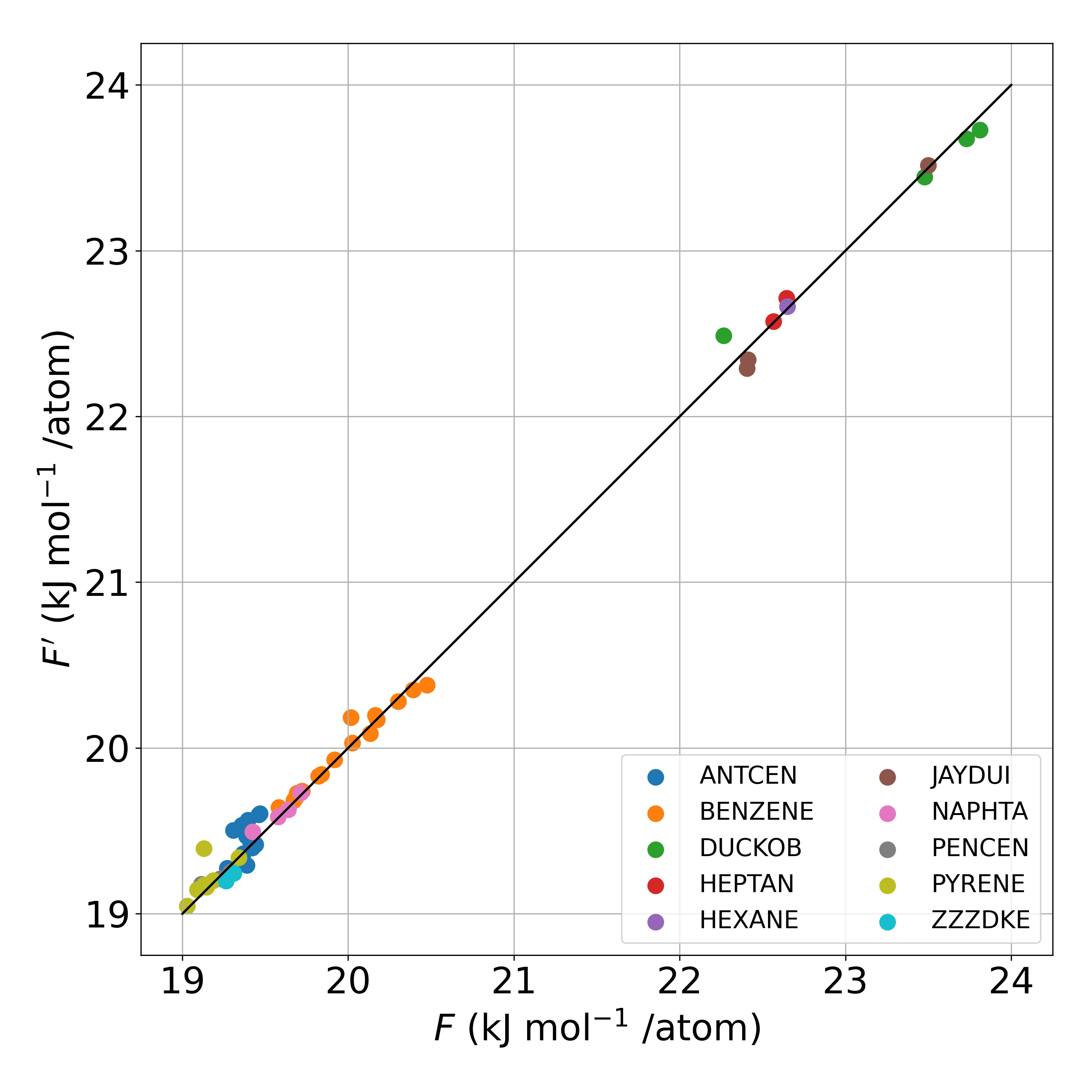}
    \caption{
        \label{fig:figure_reply}Correlation between predicted $F’$ and calculated $F$ free energies at 300 K, with a model trained on force-field structures, but with DFT free-energy predictions. Different crystal families are represented by different colors.
    }
\end{figure}

We have analyzed whether the force-field (FF) structures could be used to predict DFT free energies. In order to obtain an upper limit for the errors that such study would yield, we performed
additional calculations of DFT free energies using geometries obtained by relaxations with the AIREBO force-field.
We have used the same training (all 60 structures), validation sets and SOAP descriptors as in the  manuscript.
We performed a new optimization of the hyperparameters of the  GPR model with the same approach as presented in the manuscript and obtained values: $\sigma$=0.079, $l$=18 and $\sigma_{\epsilon}$=0.04.
The predictions were performed at 300K.
Figure \ref{fig:figure_reply} shows a correlation between predicted and calculated free energies.
We have obtained a mean absolute deviation of 0.07 kJ/mol/atom  -- almost a factor 2 higher when compared to the 0.04 kJ/mol/atom when DFT structures were used.
It is expected that once more accurate potential is used, resulting in structures closer to those obtained in DFT, a higher prediction accuracy could be achieved. This method thus shows some promise that the DFT free energies could be directly from FF structures and potentially free energies.

